\documentclass[aps,prl,twocolumn,showpacs,superscriptaddress,tightenlines,preprintnumbers,sort&compass,amsmath,amssymb,floatfix]{revtex4}

\usepackage{epsfig,graphicx,times}
\usepackage{dcolumn}
\usepackage{bm}
\usepackage{hyperref}

\begin{document}


\title{On-demand Entanglement Source with Polarization-Dependent Frequency Shift}

\author{Cheng-Xi Yang}
\affiliation{Department of Physics, Tsinghua University, Beijing 100084, China}%
\author{Yan-Bing Liu}
\affiliation{Department of Physics, Tsinghua University, Beijing 100084, China}%
\author{Xiang-Bin Wang}%
 \email{xbwang@mail.tsinghua.edu.cn}
\affiliation{Department of Physics, Tsinghua University, Beijing 100084, China}%

\date{\today}

\begin{abstract}
  We propose a polarization-dependent spatial phase
  modulation method to purify the two-photon polarization
  entanglement generated by the biexciton cascade decay in a
  single semiconductor quantum dot. In principle, our method
  can completely compensate the random phase acquired from
  the decay of the non-degenerate exciton states in time
  domain. In frequency domain, our method is equivalent to
  shifting photon frequency according to its polarization.
  The method can be applied immediately by existing
  experimental set-ups.
\end{abstract}

\maketitle \pagenumbering{arabic}

{\it Introduction.---} Quantum entanglement plays an important role
in the study of fundamental principles of quantum
mechanics\cite{sakurai}. It is also the most important resource in
quantum information
processing\cite{nielsen,decoy}.
Among all types of quantum entanglement, polarization entangled
photon-pairs are particularly useful because of easy manipulation
and transmission. There are many matured techniques to produce such
entangled pairs \emph{probabilistically}\cite{PRL_93_KiessTE,
  PhysRevLett.75.4337, Nature_04_EdamatsuK, PRL_04_FattalD},
while an \emph{on-demand} entangled photon pair is essential in many
tasks in quantum information processing.

Recently, an on-demand entangled photon-pair source was
proposed\cite{PRL_00_OliverB} and realized in a semiconductor quantum
dot system\cite{Nature_06_StevensonRM, AkopianN_PRL_06,
NatPhon_ShieldsAJ_07}. However, because of fine-structure splitting
(FSS), the relative phase of the entangled state is
randomized so that only classical correlation can be detected by
traditional time-integrated
measurement\cite{PRL_07_HudsonAJ,PRL_08_StevensonRM}. So
far, there are many methods proposed to explore this
``hidden entanglement'', for example, reducing
FSS\cite{NJP_06_YoungRJ,NJP_07_HafenbrakR,he:157405,
  PRL_09_YoungRJ}, spectral filtering\cite{AkopianN_PRL_06}, time
resolving post-selection\cite{PRL_08_StevensonRM}, and so on. Up to
now, the smallest FSS realized in experiment is about $0.3\ \mu eV$
and non-classical nature of the radiation field is verified by
directly observing violation of the Bell
inequality\cite{PRL_09_YoungRJ}. However, the entanglement quality
is considerably decreased even by very small FSS, and further
reducing FSS is very difficult in experiment. Furthermore, the
severe restriction on FSS greatly limits the selection range of
quantum dot systems. Certain quantum dots with large FSS cannot be
used even if they have distinct advantage, such as
emitting photons of frequencies in the easy transmission frequency
window in free space or optical fiber. Also, the post-selection
method in frequency domain or time domain will significantly
decrease the photon collection efficiency.

{\it Our method.---} Here we propose an experimental scheme to get
rid of \emph{all} the drawbacks listed above. In principle, our
method can \emph{completely} compensate the random phase resulted
from FSS, and therefore, greatly enlarge the selection range of
quantum dot systems. Meanwhile, our proposal just slightly reduces
the photon-pair collection rate due to the loss of the phase
modulator employed. The key point in our method is
polarization-dependent spatial phase modulation, which is
equivalent to a polarization-dependent
frequency shift operation $U(\Delta_1,\Delta_2)$ defined as
\begin{equation}
  \label{eq:3}
  \begin{split}
    & U(\Delta_1,\Delta_2) |H_1H_2;\omega_1,\omega_2\rangle
    = |H_1H_2;\omega_1,\omega_2\rangle;\\
    & U(\Delta_1,\Delta_2) |V_1V_2;\omega_1,\omega_2\rangle
    =|V_1V_2;\omega_1
      + \Delta_1, \omega_2 + \Delta_2\rangle.
  \end{split}
\end{equation}
Here $H$ and $V$ stand for horizontal and vertical polarization,
respectively, $\omega$ is the photon frequency, $\Delta$ is an
arbitrary frequency shift, and the subscripts $1$ and $2$ refer to
the first photon and second photon, respectively. After such a
unitary transformation is applied, FSS can be completely compensated, and
therefore, ``hidden entanglement'' revives.

\begin{figure}
\includegraphics{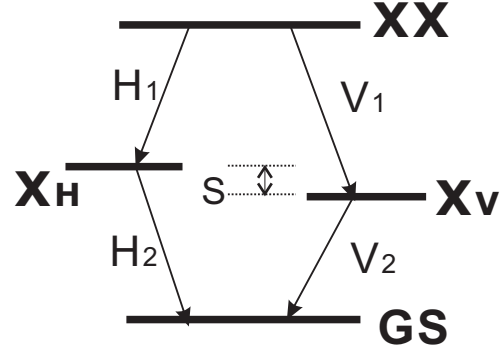}
\caption{\label{fig:levels} Energy levels of the
  semiconductor quantum dot used to generate polarization
  entangled photons. The biexciton state (XX) is a
  zero-spin state formed by two electrons and two heavy
  holes. When the dot decays, two photons are emitted
  sequentially, and their polarization is determined by the
  ``decay path''. Usually an FSS $S$ exists between the two
  excitons ($X_H$) and ($X_V$).}
\end{figure}

The energy levels of the quantum dot used for photon-pair generation
are shown in Fig.~\ref{fig:levels}. After exciting a single quantum
dot into biexciton state (XX), two photons are emitted sequentially
as the dot decays in a cascade process.
Because the two exciton states ($X_H$ and $X_V$) are not
degenerate\cite{PhysRevLett.76.3005,PhysRevB.65.195315}, the two
photons are actually entangled in the complex space of both
polarization and frequency
\begin{equation}
  \label{eq:stateFre}
  \begin{split}
  |\Psi\rangle =& \frac{1}{\sqrt{2}}
   \bigg[ \iint_{-\infty}^{\infty} d\omega_1 d\omega_2
    \Phi_H(\omega_1,\omega_2) |H_1H_2;\omega_1,\omega_2\rangle\\
   & + \iint_{-\infty}^{\infty} d\omega_1 d\omega_2
    \Phi_V(\omega_1,\omega_2) |V_1V_2;\omega_1,\omega_2\rangle
  \bigg],
  \end{split}
\end{equation}
After $U(\Delta_1,\Delta_2)$ is applied, the state of the
two photons is changed into
\begin{equation}
  \label{eq:12}
  \begin{split}
   & U(\Delta_1,\Delta_2) |\Psi\rangle\\
 = & \frac{1}{\sqrt{2}}
   \bigg[ \iint_{-\infty}^{\infty} d\omega_1 d\omega_2
    \Phi_H(\omega_1,\omega_2) |H_1H_2;\omega_1,\omega_2\rangle\\
   & + \iint_{-\infty}^{\infty} d\omega_1 d\omega_2
    \Phi_V(\omega_1-\Delta_1,\omega_2-\Delta_2) |V_1V_2;\omega_1,\omega_2\rangle
  \bigg].
  \end{split}
\end{equation}
If the two spectral functions satisfy
\begin{equation}
  \label{eq:9}
  \Phi_H(\omega_1, \omega_2) = \Phi_V( \omega_1- \Delta_1,
  \omega_2 - \Delta_2),
\end{equation}
the frequency space and polarization space are completely separated
and the two photons become maximally entangled in polarization
space:
\begin{equation}
  \label{eq:4}
  U(\Delta_1,\Delta_2) |\Psi\rangle = |\Phi^{(+)}\rangle \otimes
  \iint_{-\infty}^{\infty} d\omega_1 d\omega_2
  \Phi_H(\omega_1,\omega_2)|\omega_1,\omega_2\rangle,
\end{equation}
where $|\Phi^{(+)}\rangle = \frac{1}{\sqrt{2}} ( |H_1H_2\rangle
+ |V_1V_2\rangle )$.

The spectral functions for the two decay path of the quantum dot
system can be written as\cite{AkopianN_PRL_06, scully}
\begin{subequations}
  \label{eq:13}
  \begin{align}
    \Phi_{H}(\omega_1,\omega_2) =& \frac{\sqrt{2}\Gamma}{2\pi}
  \frac{1}{\omega_1 + \omega_2 - \omega_{0} + i\Gamma}\nonumber\\
  & \times\frac{1}{\omega_2 - \omega_{H_2} +
    i\Gamma/2},\label{eq:14} \\
  \Phi_{V}(\omega_1,\omega_2) =& \frac{\sqrt{2}\Gamma}{2\pi}
  \frac{1}{\omega_1 + \omega_2 - \omega_{0} + i\Gamma}\nonumber\\
  & \times\frac{1}{\omega_2 - \omega_{V_2} +
    i\Gamma/2}.\label{eq:20}
  \end{align}
\end{subequations}
Here, as shown in Fig.~\ref{fig:levels},
$\omega_{H_2}=\omega_{X_H}-\omega_{GS}$,
$\omega_{V_2}=\omega_{X_V} -\omega_{GS}$, and $\omega_0 =
\omega_{XX}-\omega_{GS}$, where $\hbar\omega_{XX}$,
$\hbar\omega_{X_H}$, $\hbar\omega_{X_V}$, $\hbar\omega_{GS}$
are the eigenenergy of levels $XX$, $X_H$, $X_V$, and $GS$,
respectively, and $\Gamma$ is the decay
rate of the four transitions $XX\rightarrow X_H$,
$XX\rightarrow X_V$, $X_H\rightarrow GS$, and
$X_V\rightarrow GS$\cite{AkopianN_PRL_06}. By noting
that
\begin{equation}
  \label{eq:16}
  \begin{split}
  \Phi_V(\omega_1+S, \omega_2-S) =& \frac{\sqrt{2}\Gamma}{2\pi}
  \frac{1}{\omega_1 + \omega_2 - \omega_0 + i\Gamma}\\
  & \times\frac{1}{\omega_2 -S - \omega_{V_2} +
    i\Gamma/2}\\
   =& \frac{\sqrt{2}\Gamma}{2\pi}
  \frac{1}{\omega_1 + \omega_2 - \omega_0 + i\Gamma}\\
  & \times\frac{1}{\omega_2 - \omega_{H_2} +
    i\Gamma/2}\\
   =& \Phi_H(\omega_1,\omega_2),
  \end{split}
\end{equation}
the spectral functions given in Eq.~(\ref{eq:13}) satisfy the
requirement in Eq.~(\ref{eq:9}). So after $U(-S,S)$ is applied, the
photon pair becomes maximally entangled.

Equivalently, our method can also be understood as a
polarization-dependent spatial phase modulation scheme. In
the moving reference frame in which the photons are at rest,
the radiation field can be regarded as a frozen wave
train. If we transform Eq.~(\ref{eq:stateFre}) from frequency space to
position space, the state of the field can be rewritten as
\begin{equation}
  \label{eq:evo}
  \begin{split}
  |\Psi\rangle =& \frac{\sqrt{2}\Gamma}{c} \iint_{0>x_1>x_2}
  \frac{1}{\sqrt{2}}\left( |H_1H_2\rangle +
    e^{iS(x_1-x_2)/(\hbar c)} |V_1V_2\rangle\right)\\
  & \times e^{\frac{\Gamma}{2}(x_2+x_1)/c} e^{i(\omega_{H_1} x_1 +
    \omega_{H_2} x_2)/c} |x_1,x_2\rangle dx_1 dx_2,
  \end{split}
\end{equation}
where $x_1$ and $x_2$ refer to the position of the first photon and
second photon, respectively. Eq.~\eqref{eq:evo} is equivalent to the
result given in Ref.~\cite{PRL_07_HudsonAJ}.

Obviously, if we take the polarization-dependent spatial phase
modulation $U$ as
\begin{equation}
  \label{eq:30}
  \begin{split}
    & U |H_1H_2;x_1,x_2\rangle
    = |H_1H_2;x_1,x_2\rangle;\\
    & U |V_1V_2;x_1,x_2\rangle
    = e^{-iS(x_1-x_2)/(\hbar c)} |V_1V_2;x_1,x_2\rangle,
  \end{split}
\end{equation}
the polarization space and position space are completely separated.
To realize the second line of the transformation above, we simply take
the following separate operation for each photon
\begin{equation}
\begin{split}
&|V_1,x_1\rangle \rightarrow e^{-iSx_1/(\hbar c)} |V_1,x_1\rangle \\
&|V_2,x_2\rangle \rightarrow e^{iSx_2/(\hbar c)} |V_2,x_2\rangle
\end{split}
\end{equation}
This is equivalent to applying the following separate phase
modulation to vertical polarized mode for each photon
\begin{equation}
  \label{eq:10}
  \varphi_1(x_1)=-\frac{S x_1}{\hbar c} \qquad
  \varphi_2(x_2)=\frac{S x_2}{\hbar c},
\end{equation}
By noting that the photons are moving with a constant
velocity $c$, the spatial phase modulation
in Eq.~(\ref{eq:10}) is equivalent to linearly changing the
compensating phase at a fixed point as
\begin{equation}
  \label{eq:2}
  \varphi_1(t)=St/\hbar \qquad \varphi_2(t)=-St/\hbar.
\end{equation}
This is just the frequency shift by $-S$ and $S$, respectively, to each photon.

\begin{figure}
  \includegraphics{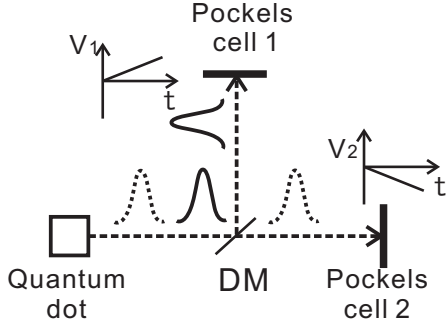}
  \caption{\label{fig:setup} Experimental set-up for our
    proposal. The first photon and second photon are separated by a
    dichroic mirror (DM). The two Pockels cells start to run before
    the photons arrive. They make reverse phase modulation.
  }
\end{figure}

\begin{figure}
  \includegraphics[width=8cm]{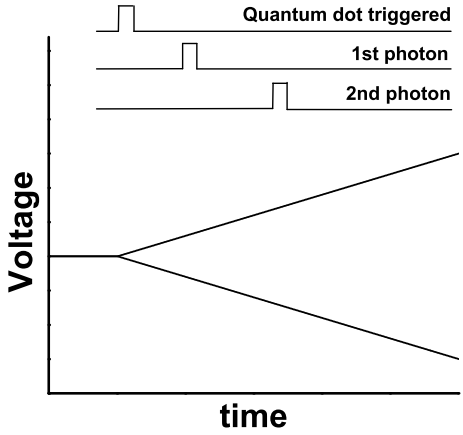}
  \caption{\label{fig:Torder} Scan voltage v.s. time.
    Compensating phase starts to increase (decrease) for
    photon 1 (2) according to Eq.~(\ref{eq:2}).
  }
\end{figure}

{\it Realization---} The frequency shift (spatial phase modulation) can be
done by using two polarization-dependent phase modulators for the
first photon and second photon, respectively, as shown in
Fig.~\ref{fig:setup}. Similar phase modulation has been
used in laser spectroscopy, such as Pound-Drever-Hall laser
frequency stabilization\cite{black:79}. The modulator for
the first (second) photon starts
to linearly increase (decrease) the compensating phase
(Fig.~\ref{fig:Torder}) before the first (second) photon
arrives. The later the first (second) photon arrives
at modulator $1$ ($2$), the higher (lower) the compensating
phase is
applied. If compensating phases are modulated according to
Eq.~\eqref{eq:2}, the random phase in Eq.~(\ref{eq:evo}) can be
completely removed, provided that the duration of phase modulation
covers the duration of the field radiation.

Technically, such phase modulation can be accomplished by a
commercially available optical device, such as a Pockels cell, which
introduces a phase shift to vertically polarized mode. The
phase shift is proportional to the scan voltage $V(t)$, i.e.
$\varphi(t)=\alpha V(t)$, where $\alpha$ is the phase sensitivity of
the Pockels cell. By linearly changing $V(t)$, Eq.~\eqref{eq:2} can
be easily satisfied. In particular, the relationship between voltage
change rate $dV/dt$ and frequency split $S$ reads
\begin{equation}
  \label{eq:6}
  \frac{dV}{dt}=\frac{S}{\hbar\alpha}.
\end{equation}
As far as we have known, the phase sensitivity $\alpha$ of
commercially available Pockels cells can be up to $52\ mrad/volt\ @\
830 nm$\cite{conoptics}. This means that a rising rate at $30\ V/ns$
is required for removing an FSS of $1\ \mu eV$. To compensate larger
FSS, we can arrange several Pockels cells in series along one photon's
path. Furthermore, because the duration of the field
radiation is about several
nanoseconds\cite{PRL_08_StevensonRM}, the scan voltage
needs only last several nanoseconds. Therefore, the maximal voltage
requested is only a few hundred volts according to Eq.~\eqref{eq:6},
and this is easily accessible.

\begin{figure}
  \includegraphics[width=8cm]{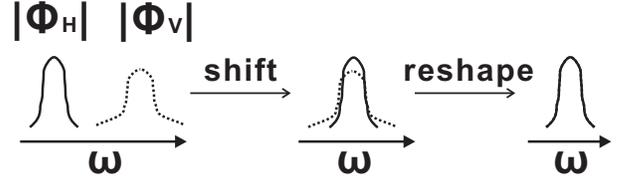}
  \caption{\label{fig:shift} Operation to maximize the
    overlap of $|\Phi_H|$ and $|\Phi_H|$. In the right
    figure, $|\Phi_H|$ and $|\Phi_H|$
    overlap perfectly.
  }
\end{figure}

{\it Discussion---} Our method is not only useful in cascading decay
process in quantum dot system, but also useful in many cases which
need to increase the frequency indistinguishability, such
as parametric-down-conversion process. Consider two spectral
functions $\Phi_H = |\Phi_H|e^{i\varphi_H}$ and $\Phi_V = |\Phi_V|
e^{i\varphi_V}$, where $\Phi_H$, $\Phi_V$, $\varphi_H$, and
$\varphi_V$ are functions of frequency. The larger $|\Phi_H|$ and
$|\Phi_V|$ overlap, the higher the indistinguishability is. As shown
in Fig.~\ref{fig:shift}, by applying our frequency shift method, the
overlap can be significantly increased. After that, we may further
increase the overlap by changing the shape of the two spectral
functions through frequency shift: Replacing $\Delta_1$ and
$\Delta_2$ by certain appropriate functions of $\Delta_1(\omega_1)$
and $\Delta_2(\omega_2)$ in Eq.~(\ref{eq:3}), $|\Phi_H|$ and
$|\Phi_V|$ will be reshaped into (almost) the same form. Meanwhile,
the frequency dependent terms $e^{i\varphi_H}$ and $e^{i\varphi_V}$
can be removed by the method given in Ref.~\cite{guo}. So in
general, after this three-step operation, one may obtain (almost)
perfect indistinguishability. (Fortunately, for the spectral
functions given in Eq.~(\ref{eq:13}), the frequency shift alone is
enough.)

Very recently, Guo's group presented an experimental scheme to
improve the quality of entangled photon pairs generated by a
quantum dot with FSS\cite{guo}. Here, we would like to compare our work
with theirs. Although both works aim at improving the quality of
entangled photon pairs, they are different in the main idea, the
result, the method, and the realization. In particular, we use the
frequency shift method and as a result, we can in principle obtain perfect
entangled photon pairs given the spectral functions of
Eq.~(\ref{eq:13}) which satisfy Eq.~(\ref{eq:9}). In their
work\cite{guo}, there is no frequency shift. Therefore, it is in
principle not possible to obtain perfect overlap of two spectral
functions if they initially have different frequency centers.
Explicitly, their scheme is to change $e^{i\varphi_H}$ and
$e^{i\varphi_V}$ in the spectral functions $|\Phi_H|e^{i\varphi_H}$
and $|\Phi_V| e^{i\varphi_V}$. Hence, through their scheme, the
maximum achievable overlap of two spectral functions can never
exceed the amount of overlap of $|\Phi_H|$ and $|\Phi_V|$, while our
scheme can produce perfect overlap between any two spectral
functions if they satisfy Eq.~(\ref{eq:9}). As was just discussed
above, in cases that Eq.~(\ref{eq:9}) is not satisfied, our method
can still help to obtain almost perfect indistinguishability if
combined with some further manipulations.

In conclusion, we have proposed an experimental scheme to purify the
entanglement of photon-pairs produced by a semiconductor quantum
dot. The entanglement distortion caused by FSS can, in principle, be
fully corrected. Since post-selection techniques are not required,
our method does not decrease the photon pair collection efficiency.

\begin{acknowledgments}
{\it Acknowledgments---} We would like to thank Jia-Zhong Hu and
Ming Gao for helpful discussions. This work was supported in part by
the National Basic Research Program of China grant nos 2007CB907900
and 2007CB807901, NSFC grant number 60725416, and China Hi-Tech
program grant no. 2006AA01Z420.
\end{acknowledgments}


\end{document}